\documentclass[12pt]{article}

\usepackage{graphicx}
\usepackage{amssymb}
\usepackage{amsmath}
\usepackage{hyperref}

\begin{document}



\title{Correlation and Autocorrelation of Data on Complex Networks}



\author{Rudy Arthur\\
{University of Exeter, Department of Computer Science,}\\
{Stocker Rd, Exeter EX4 4PY}\\
{E-mail:  r.arthur@exeter.ac.uk}
}
\maketitle

\begin{abstract}
    Networks where each node has one or more associated numerical values are common in applications. This work studies how summary statistics used for the analysis of spatial data can be applied to non-spatial networks for the purposes of exploratory data analysis. We focus primarily on Moran-type statistics and discuss measures of global autocorrelation, local autocorrelation and global correlation. We introduce null models based on fixing edges and permuting the data or fixing the data and permuting the edges. We demonstrate the use of these statistics on real and synthetic node-valued networks.
\end{abstract}


\section{Introduction}\label{sec:introduction}

This paper studies networks where some numerical data is observed at every node. For example, diffusion processes \cite{lopez2008diffusion}, representing the spread of rumors \cite{de2018fundamentals}, information \cite{zhang2016dynamics}, disease \cite{pastor2015epidemic} and so on, are often studied using the SIR model and quantities like activation potentials or probability of infection have values at every node. Other representative examples of node data include: population counts \cite{expert2011uncovering} or destination `attractiveness' \cite{moya2018dynamic} when nodes represent locations; number of posts \cite{arthur2019human} or their sentiment\cite{tan2011user,fornacciari2015social,pozzi2016sentiment}, when nodes represent users in a social network or more abstract measures, like relevance scores of skills for jobs \cite{alabdulkareem2018unpacking}.

In the network science literature these \emph{node-valued} networks have received somewhat less attention than \emph{node-attributed} networks. In particular, the use of node attributes to inform or improve community detection algorithms has been extensively studied, see the reviews \cite{bothorel2015clustering, chunaev2020community}. Node attributes are generally categorical. For example, the famous US political blog network \cite{adamic2005political} has a binary (conservative/liberal) attribute at each node. Citation networks often have an attribute which labels the topic or words used in the text corresponding to each node \cite{sen2008collective, yang2015network}. This work concerns networks with real numbers associated with each node. I will refer to these as \emph{node-valued} networks to distinguish them from  \emph{node-attributed} networks, however there is a close link between the two. For example, by applying topic modelling, \cite{jia2017node} associates a 10-dimensional vector with real number components to each node. One could use each component of the vector as a node-value or one could take the topic with the largest component as the node attribute.

More formally, we have network with $N$ nodes and $E$ edges together with an $N$ dimensional vector $\mathbf{x}$ which has the value $x_i$ on  node $i$. The aim of this work is to use simple summary statistics to characterise the relationship between `nearby' values of $\mathbf{x}$ - the autocorrelation - as well as the relationship between two vectors, $\mathbf{x}$ and $\mathbf{y}$, representing different variables defined on the same network - the correlation. The difficulty will be to  properly account for and incorporate the underlying network structure into the summary statistic and its distribution.

Along the same lines as the above, Coscia \cite{coscia2021pearson} defines a correlation measure that incorporates information about the network into the correlation score between two variables measured at the nodes (see also \cite{devriendt2022variance}). \cite{coscia2021pearson} notes the similarity between their correlation measure and Moran's $I$ \cite{moran1950notes}. Moran's $I$ is a measure of spatial autocorrelation. It quantifies the extent to which similar values of $\mathbf{x}$ are close to each other. Moran's $I$ is given by
\begin{equation}\label{eqn:moran}
    I = \frac{N}{|W|} \frac{ \sum_{ij}^N (x_i - \bar{x}) w_{ij} (x_j - \bar{x}) }{ \sum_i^N (x_i - \bar{x})^2}
\end{equation}
Where $\bar{x}$ is the mean of $\mathbf{x}$. The  weight matrix $W$ has elements $w_{ij}$ and $|W| = \sum_{ij} w_{ij}$. In spatial analyses the weight matrix is often the row-normalised adjacency matrix of the network constructed by taking the regions $i$ as nodes and adding edge $ij$ if $j$ borders $i$ \cite{dale2014spatial}. Alternatively, as in 
 \cite{coscia2021pearson}, a distance decay function is sometimes used e.g. $w_{ij} = e^{-c d_{ij}}$ for some constant $c$ where $d_{ij}$ is the spatial distance between regions.

Here, we simply note that spatial statistics, like Moran's $I$ and others to be discussed later, can be computed for any network. There is nothing in Equation \ref{eqn:moran} requiring $W$ to be planar. Furthermore, $I$, and the other spatial statistics to be discussed, impose no strong restrictions, like positive definiteness \cite{coscia2024pearson}, on the interaction matrix $W$. As long as it provides a reasonable description of the relationships between the nodes we can use it. In particular, the adjacency matrix or a distance decay matrix like $e^{-cd_{ij}}$ should work and the value of $I$ will, as in the spatial case, reflect the degree to which similar values of $\mathbf{x}$ are close to each other, where close means \emph{connected by short paths}. While some networks may be quite similar to spatial maps, for example transport networks, other types, like social networks, can have small world effects \cite{watts1998collective}, modular structure \cite{newman2006modularity} and scale free degree distributions \cite{newman2001random} which spatial networks usually lack. Understanding how to interpret Moran's $I$ and other spatial statistics on non-spatial networks is the aim of this work.

We will summarise distributions using one-sided p-values - the probability that the observed value or greater would be observed by chance under a null model. P-values can be estimated by randomly permuting the values on the network and comparing the observed statistics against the distribution of the statistic evaluated on permuted networks. This is the standard approach in spatial statistics \cite{rey2009pysal} and we will refer to this as the `data-permutation' null. When the nodes represent physical locations it makes sense to view the underlying network as fixed and vary the measurements. When we have a non-spatial network it is also possible for the links to vary, as in a social or contact network. In these cases it might be more appropriate to consider a null model where the node values are fixed and the network edges are varied. While there are many possible null models we could use, since the degree distribution is usually a key factor in explaining network structure \cite{newman2001random} we will use the configuration model to construct an ensemble of networks with the same degree-distribution as the observed network to generate a test distribution for the `configuration' null.

 We will first study global autocorrelation using Moran's $I$, comparing the different null models described above in Section \ref{sec:global}. We then introduce `Local Indicators of Spatial Association' (LISA) statistics \cite{anselin1995local} for finding clusters of interesting values and generalise these to `Node Indicators of Network Association' (NINA) statistics in Section \ref{sec:local}. We discuss correlation of two different variables observed on the same network and generalise a definition of network correlation, Lee's $L$ \cite{lee2001developing}, to arbitrary networks in Section \ref{sec:bivariate}. We then introduce the idea of distance class and the correlogram in Section \ref{sec:correlogram} in order to study how clustering behaviour changes as we consider next-nearest neighbours and beyond. Finally, in Section \ref{sec:wikipedia} we analyse a small but realistic network constructed from Wikipedia articles, using the tools we develop to perform an exploratory data analysis of page views and other measurements.
 
\section{Global Autocorrelation}\label{sec:global}

\subsection{Definition}\label{sec:globaldef}

Although many other measures of spatial autocorrelation exist \cite{cliff1981spatial} we will primarily focus on Moran's $I$. This measure can be better understood by rewriting Equation \ref{eqn:moran}. First let
$$
z_i = (x_i - \bar{x})
$$
then, borrowing terminology from spatial data analysis, define the (right) lagged-vector as
$$
\tilde{z}_i = \sum_j^N w_{ij} z_j
$$
Equation \ref{eqn:moran} can be written in vector form as
\begin{equation}
I = \frac{N}{|W|} \frac{ \mathbf{z} \cdot \mathbf{ \tilde{z} } }{ | \mathbf{z} |^2 }
\end{equation}
For the weight matrix $W$ we will mostly use the row normalised adjacency matrix $A$. In this case $\sum_j A_{ij} = 1$ for every $j$ and therefore $|W| = |A| = N$, cancelling out the constant factor in front. 

When $W$ is the row-normalised adjacency matrix, $\tilde{z}_i$ has a simple interpretation as the average value of $\mathbf{z}$ over the neighbours of $i$. Therefore, Moran's $I$ is a global comparison of values observed at the nodes with the average value of each node's neighbours. For a row normalised weight matrix, Moran's $I$ generally gives values in the range $[-1,1]$, the max and min are set by the largest eigenvalues of $W$. The row normalisation makes $W$ a stochastic matrix so the largest eigenvalue has absolute value 1, though there are some cases, e.g. isolated nodes, which result in values outside this range, see \cite{de1984extreme} for more details. Generally, $I$ will be larger when similar values of $z_i$ are frequently connected via $W$ and close to zero for randomly distributed data \cite{cliff1981spatial}. To distinguish the quantity evaluated on an arbitrary network from the spatial measure we will refer to this number as the \emph{network Moran Index}.

A similar quantity to $I$ is the assortativity coefficient \cite{newman2003mixing},
\begin{equation}\label{eqn:assortativity}
    \frac{ \sum_{x,x'} xx'(e_{xx'} - a_x b_{x'})}{\sigma_a \sigma_b}
\end{equation}
The sums are over all observed values; $e_{xx'}$ is the fraction of edges that join values $x$ and $x'$; $a_x$ and $b_{x'}$ are the row and column sums of $e_{xx'}$ and $\sigma_a, \sigma_b$ are the standard deviations of the distributions $a_x$ and $b_{x'}$. There are also local versions of assortativity defined in the literature \cite{piraveenan2008local} which we will not discuss in detail. The assortativity is usually computed for discrete data and real valued data is frequently binned in order to calculate $e_{xx'}$ \cite{hagberg2008exploring}. If data is not binned $e_{xx'}$ will likely equal the adjacency matrix, since all the values observed will likely be distinct. In this case Equation \ref{eqn:assortativity} becomes
\begin{equation}
  \frac{ \sum_{i,j} x_i A_{ij} x_j - x_i \frac{k_i k_j}{2E} x_j}{\sigma_k^2}
\end{equation}
where $\sigma_k^2$ is the variance of the node degree and $2E = \sum_{ij} A_{ij}$. The first term in the numerator is similar to the numerator of the Moran index, apart from the mean subtraction. In fact this is the numerator of another closely related spatial statistic called the Getis-Ord statistic \cite{getis1992analysis}. Just like the Moran Index, $x_i \sum_j A_{ij} x_j$ compares the value at the node with the average value at the neighbours. In contrast to Getis-Ord and Moran statistics, the assortativity coefficient is normalised by the variance of the node degree rather than the variance of the data, reflecting an emphasis on the connectivity of the network.

The second term in the numerator has the adjacency matrix replaced by $k_i k_j / 2E$, which is proportional to the expected number of connections between $i$ and $j$, fixing the node degrees. This is form is similar to the modularity function \cite{newman2006modularity} used in community detection, but with each edge's contribution weighted by the values of the nodes it links instead of a community indicator function. The expected value of $A_{ij}$ under the configuration null is $k_i k_j / 2E$, so assortativity values close to zero imply no clustering of $\mathbf{x}$  beyond what would be expected given the node degrees.

While assortativity has been widely used to assess similarity of node values, $I$ offers some advantages. First, it uses centred data and is normalised by the variance of the data rather than the variance of the node degree, which makes it invariant to rescaling and shifts in the values of $\mathbf{x}$. It is flexible with respect to the definition of neighbour, so alternative weight matrices like $w_{ij} = e^{-d_{ij}}$ or $w_{ij} = A_{ij}^k$ can be used, as we will see in Section \ref{sec:correlogram}. Unlike assortativity, the Moran index does not incorporate the null model into its definition, giving more flexibility in how the null is specified, in particular we can apply the `data-permutation' null from spatial statistics. Lastly, from the definition of $I$ it is apparent that the index can be computed straightforwardly, without requiring any rounding or binning.

\subsection{Examples}

We will show examples on small networks for clarity. $I$ is quite fast to compute, the main cost being the sparse matrix-multiplication required to obtain $\mathbf{\tilde{z}}$. Randomising the data or the network hundreds of times to compute a testing distribution can be time consuming, though for the relatively small networks considered here only takes a few seconds on a single core. To build intuition we generate autocorrelated data on a network using the value propagation algorithm described in \cite{coscia2021pearson}:
\begin{itemize}
    \item Choose a node to be the source and assign the value $x_{source} = 1$, with the other values set to $x_i = 0$ initially.
    \item Update the values of $x_i \rightarrow \frac{1}{k_i} \sum_{j \in nbr(i)} x_j$. Where $k_i$ is the degree of $i$ and $nbr(i)$ is the set of nodes that are connected to $i$. This is equivalent to $x_i \rightarrow  \sum_{j} A_{ij} x_j$. 
    \item The source node value is fixed at 1.
    \item Repeat this $M$ times.
    \item Add a random number to each of the $x_i$ where the random number is drawn from a Gaussian distribution with mean $0$ and variance $\sigma^2$.
\end{itemize}

\begin{figure}
    \centering
    \includegraphics[width=\textwidth]{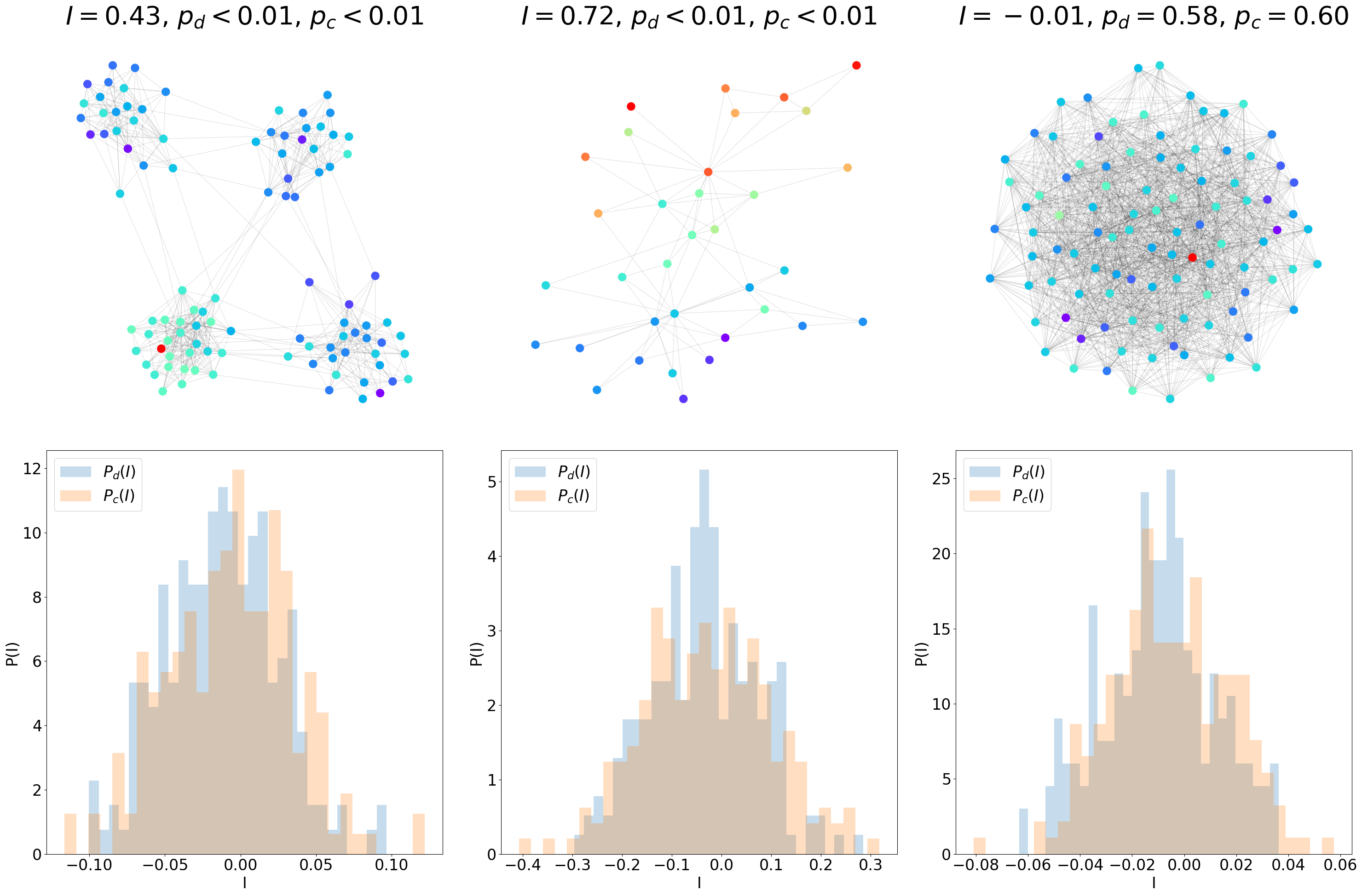}
    \caption{Left: LFR benchmark network, Middle: Karate club, Right: Erd\H{o}s-Renyi graph. Top: Force directed layout, node colours indicate node values initialised as described in the text with $M=10$ for LFR and Karate club and $M=30$ for ER and  $\sigma=0.1$ for all three. Bottom: Distribution of values from random permutations of the data, and configuration model.}
    \label{fig:gloabl_moran}
\end{figure}
Figure \ref{fig:gloabl_moran} shows three networks: an LFR benchmark network \cite{lancichinetti2008benchmark}, the karate club network \cite{zachary1977information} and an Erd\H{o}s-Renyi graph \cite{erdds1959random}. The node data is generated by the value propagation process described above. This and subsequent network values are coloured such that high values are red, low values are blue. The LFR and karate club networks have obvious community structure where the value propagation happens readily inside the source community, but values do not easily propagate outside. This results in positively autocorrelated data: nodes in the source community have high values, nodes outside the source community have low values. We compute significance under the two null models discussed. Randomly permuting the data gives the distribution $P_d(I)$ and p-value $p_d$. The configuration model, randomly swapping the edges of the network while keeping the node degrees and values, $x_i$, fixed  gives $P_c(I)$ and p-value $p_c$. The p-value is vanishingly small under both nulls and the network Moran Index is relatively large for both LFR and karate club networks, indicating significant positive autocorrelation.

The same process on the ER network does not result in a large or significant autocorrelation under either null model. When a large fraction of the nodes in a network are connected, this results in low autocorrelation. This is both because the value propagation algorithm has to average over a large number of low value nodes and is also an intrinsic property of $I$. In the limit of a clique, using the row normalised adjacency matrix as the weight matrix $A_{i \neq j} = \frac{1}{N-1}$ implies
\[
\sum_{ij}^N z_i A_{ij} z_j = \frac{1}{N-1} \left( \sum_{i}^N z_i \right)^2
\] 
so $I = \frac{1}{N-1}$, which is small for large N.  From this we infer that when the network is very densely connected, $I$ may not be a useful metric. Simply, paths are so short that everything is a near neighbour of everything else and the sum $\sum_j A_{ij} z_j$ averages over a large fraction of the network for every $i$, resulting in little variation across the nodes.

However, sparse networks with modular structure are common in practice. In this case $I$ can be a useful indication of data autocorrelation, or clustering of similar values. An interesting question is how individual nodes contribute to the autocorrelation. A node can make a large contribution to $I$ due to a large $z_i$ value at the node \emph{or} by that node having many neighbours making contributions with the same sign, i.e. large $\tilde{z}_i = \sum_j A_{ij} z_j$. This brings us to the idea of a `local indicator', which we discuss in the next section.

Before turning to this we briefly note that the principle of homophily says that nearby nodes will influence each other to have similar values, which suggests that positive autocorrelation is more likely in practice. However negative autocorrelation is possible. For example, a bipartite network with all the values $x_i=1$ in one part and $x_i=0$ in the other will result in a large, negative, value of $I$.

\section{Local Indicators}\label{sec:local}

\subsection{Definition}

Anselin \cite{anselin1995local} introduced the idea of a `Local Indicator of Spatial Association' (LISA) statistic. These are local functions of the data 
$\Lambda_i = f(x_i, \{x_j, j \in nbr(i)\})$ that are designed so that the sum of local indicators gives a global indicator $\Lambda$ up to a constant: $\sum_i \Lambda_i \propto \Lambda$. The prototypical example is the local Moran statistic
\begin{equation}\label{eqn:localI}
    I_i = \frac{z_i}{\sum_j z_i^2} \sum_j w_{ij} z_j =  \frac{z_i \tilde{z}_i }{ | \mathbf{z} |^2 }
\end{equation}
 Up to scaling, $I_i$ is simply the product of the observed and lagged data at $i$ and $\sum_i I_i \propto I$. Again, we can compute this on any network and assess significance against null models where the data is varied or the network is varied. To vary the network we will use the configuration model as before. To vary the data, as in \cite{anselin1995local} we use \emph{conditional} randomisation: to compute the significance of $I_i$, $z_i$ is fixed while data at the other nodes is permuted. To distinguish the network version from the spatial version we refer to `Node Indicators of Network Association' (NINA) statistics and the statistics above as the \emph{node Moran Indices.}

An assessment of the local/node Moran index is often paired with a Moran scatter-plot \cite{anselin2019moran}, a plot of $z_i$ against $\tilde{z}_i$. The slope of this line is equivalent to the Moran index. Examining the four quadrants of the plot (high-high, high-low, low-high, low-low) gives information about areas of the map, or network, which are positively and negatively autocorrelated and can be very useful in interpreting local indices. 

\subsection{Examples}

\begin{figure}
    \centering
    \includegraphics[width=\textwidth]{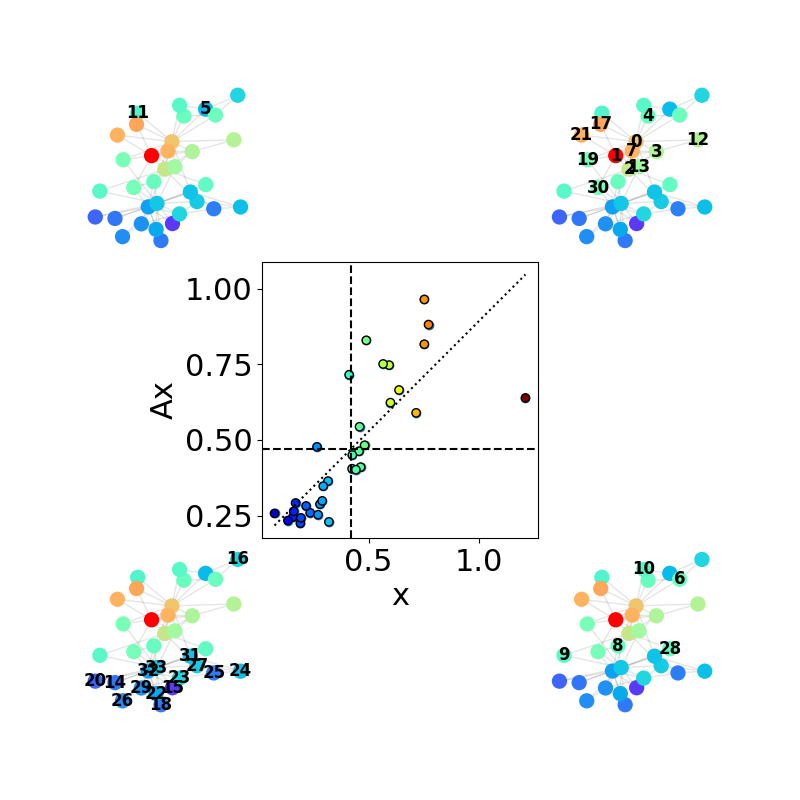}
    \caption{Center shows the Moran scatter plot. The nodes in each quadrant are labelled in the corresponding network diagram.}
    \label{fig:moran_scatter}
\end{figure}
 Figure \ref{fig:moran_scatter} shows the karate club network where the propagation process starts from node labelled 1 and is run for $M=10$ iterations. Node 1 is directly connected to the leader of one `faction' but not the other. The upper right quadrant shows the high-high associations i.e. high $x$ values that are neighbours of high $x$ values. Nodes in this quadrant are close neighbours of the source. The low-low quadrant identifies the other faction. These nodes are in the other community and so values do not easily propagate from the source to these nodes. The off diagonal quadrants are also interesting, they identify low nodes which are neighbours of high nodes (upper left) and high nodes which are neighbours of low ones (bottom right).  In this case, the bottom right nodes represent a kind of transitional zone across which values change from high to low.

\begin{figure}
    \centering
    \includegraphics[width=\textwidth]{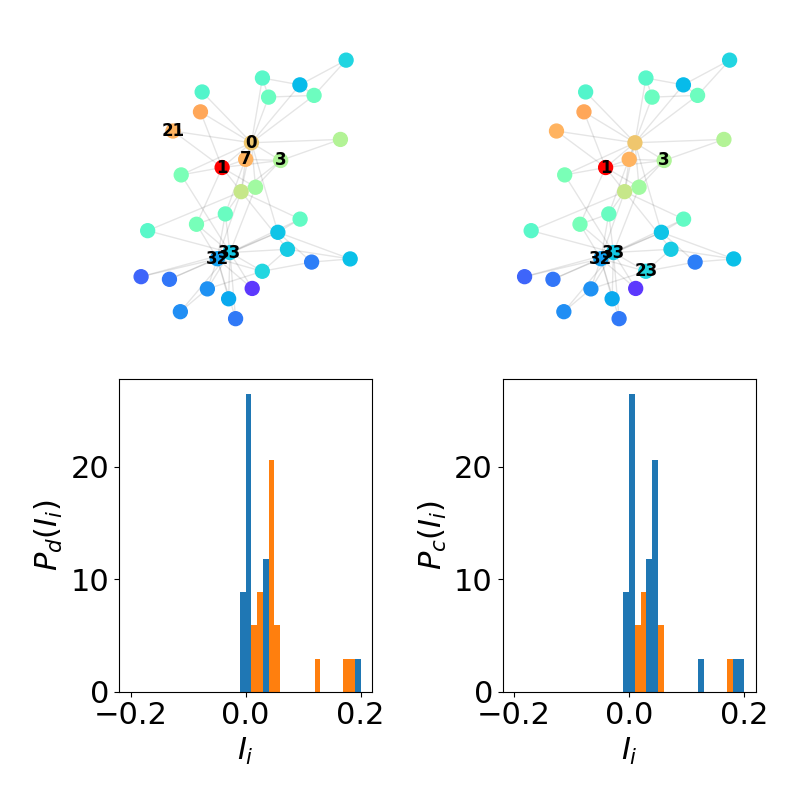}
    \caption{Local network statistics. Left: Data-permutation null. Right: Configuration model null. Top: Network with `interesting' nodes under the corresponding null models labelled. Bottom: Histograms of node Moran index, bins containing significant values are coloured orange. }
    \label{fig:local_moran}
\end{figure}
Figure \ref{fig:local_moran} shows a histogram of the node Moran index, where `significant' values ($p < 0.01$) are identified using either conditional randomisation of the data or the configuration model. A similar but non-identical set of nodes are identified under both nulls. Note that it is not simply high values of $I_i$ that are significant. The clusters are a small group of nodes around the source which have unusually high values, and a number of nodes close to the leader of the other faction with unusually low values. In general the node Moran index is useful to find these small, dense clusters, as we will see in Section \ref{sec:wikipedia}.

Because we test every node, using a naive significance level $\alpha$ runs into the problem of multiple comparisons. Standard approaches are selecting a very low significance threshold (as here), Bonferroni correction or False Discovery Rate, as described in \cite{anselin1995local}. Following best practice in spatial analysis \cite{anselin1995local,anselin2019local}, we do not recommend uncritical application of any simple threshold for significance. Rather, values of $I_i$ which are unusual under the null suggest subsets of \emph{interesting} nodes \cite{efron2021computer}. Thus any reasonably low threshold can be used to identify nodes which should be investigated in more detail by e.g. looking at other node statistics (degree, centrality etc.); using other node metadata or looking at nodes in the context of mesoscopic structure (community, core-periphery etc.).

\section{Bivariate Correlation}\label{sec:bivariate}

\subsection{Definition}

Coscia in \cite{coscia2021pearson} develops a measure of association between two different variables measured on the nodes of the same network. For simplicity assume the data $\mathbf{x}$ and $\mathbf{y}$ have had their mean subtracted, then the definition in \cite{coscia2021pearson} is
\begin{equation}\label{eqn:coscia}
    \rho_G(x, y) = \frac{\sum_{ij} w_{ij} x_i y_j}{ \sqrt{ \sum_{ij} w_{ij} x_i x_j} \sqrt{ \sum_{ij} w_{ij} y_i y_j} }
\end{equation}
Problems arise from the denominator, namely, for arbitrary $W$ the terms under the square root need not be positive \cite{coscia2024pearson}. Indeed, using the notation of Section \ref{sec:globaldef}, the term under the square root is
\begin{equation}\label{eqn:netvar}
   \sum_{ij} w_{ij} x_i x_j = \mathbf{x} \cdot \mathbf{ \tilde{x} } 
\end{equation}
which is the the numerator of $I$ and can certainly be 0 or negative.

Another problem not discussed by \cite{coscia2024pearson} but similar to issues discussed by \cite{lee2001developing}, is the fact that there is only one factor of $W$ in the numerator of Equation \ref{eqn:coscia}. Summing over the $j$ index, the correlation is a dot product between the data $x$ and the lagged data $y$. Thus only the $y$ data `sees' the network in the comparison. This lack of symmetry is potentially problematic if, for example, $x$ is strongly autocorrelated but $y$ is not. Using the `network variance', equation \ref{eqn:netvar}, in the denominator could mitigate this but introduces the problems with positive definiteness mentioned.

Lee \cite{lee2001developing} provides an alternative approach. Define
\begin{equation}\label{eqn:lee}
L = \frac{N}{\sum_i \left( \sum_j w_{ij} \right)^2 } \frac{ \tilde{x} \cdot \tilde{y} }{|x||y|}
\end{equation}
For a row normalised $W$ the constant factor in front disappears. $L$ uses the magnitude of the data in the denominator, which is a sum of squares so there is no issue with negative roots. $L$ is a normalised dot product between the lagged $x$ and lagged $y$ data, so both datasets are influenced by the network structure. It is worth noting that if the usual weight/adjacency matrix is used, where $A_{ii} = 0$, then the value of $x_i$ interacts with the value at $y_i$ only through the neighbours of $i$. To account for this we could add self loops or use a distance matrix and kernel function like $w_{ij} = e^{-d_{ij}}$. For simplicity when computing $L$ we set $A_{ii} = 1$ so that the values at $x_i$ and $y_i$ interact directly in the $i^{th}$ term of the sum.

\subsection{Examples}

\begin{figure}
    \centering
    \includegraphics[width=\textwidth]{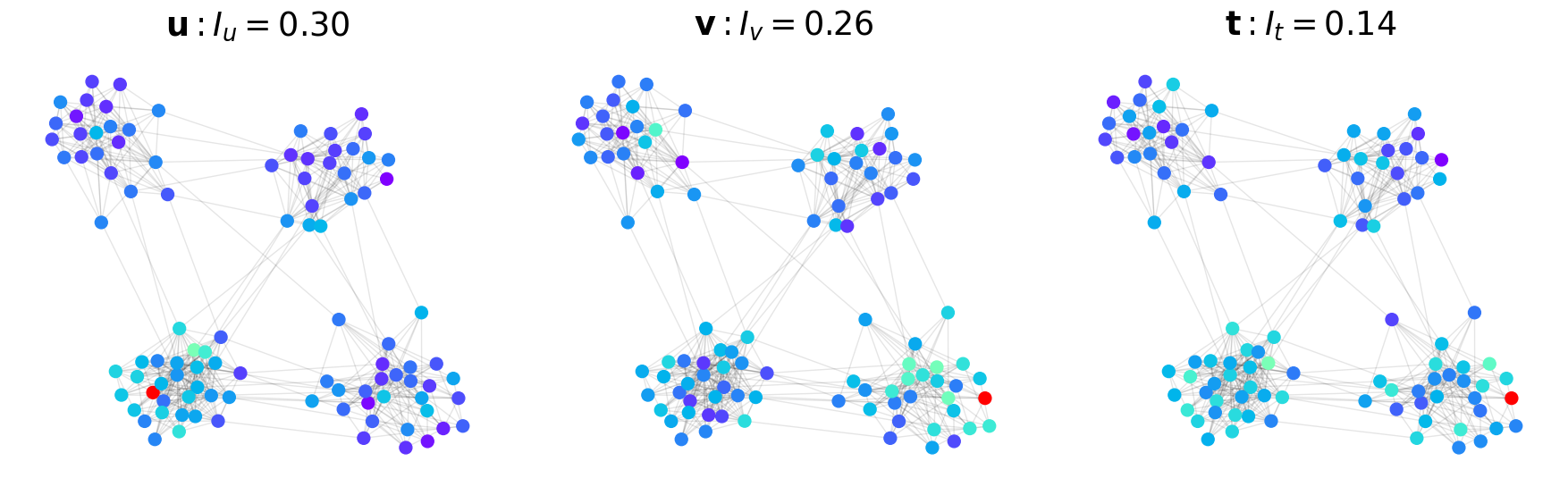}
    \caption{Different propagation processes on the same LFR network. The network Moran index is shown for each data set. All $I$ values are highly statistically significant.  }
    \label{fig:lee_statistic}
\end{figure}

\begin{table}
    \centering
    \begin{tabular}{|c||c|c||c|c|c|}
    \hline
          & $\rho_{pearson}$ & $p_{pearson}$ & $L$ & $p_d$ & $p_c$\\ \hline
$\mathbf{u},\mathbf{v}$ & -0.04 & $ 0.66$ & -0.03 & $0.96$ & $ 0.99$ \\ \hline
$\mathbf{u},\mathbf{t}$ & 0.10 & $0.13$ & \textbf{0.14} & $< 0.01$ & $< 0.01$ \\ \hline
$\mathbf{v},\mathbf{t}$ & \textbf{0.49} & $< 0.01$ & \textbf{0.13} & $< 0.01$ & $< 0.01$ \\ \hline
    \end{tabular}
    \caption{Pearson and Lee correlation for data shown in Figure \ref{fig:lee_statistic}. Interesting observations (low p-values) are highlighted in bold.}
    \label{tab:lee_statistic}
\end{table}

Figure \ref{fig:lee_statistic} shows three data sets defined on the same LFR network with 100 nodes. The $\mathbf{u}$ and $\mathbf{v}$ data are generated by running the value propagation algorithm starting at different source nodes, $M=5$ and $\sigma=0.1$. The $\mathbf{t}$ data starts with the same source as $\mathbf{u}$, runs for $M=5$ steps, swaps the source to the same position as $\mathbf{v}$, runs for another $M=5$ steps and then adds Gaussian noise with $\sigma=0.1$. This produces three weakly but significantly autocorrelated datasets. Their Pearson and Lee correlations are shown in Table \ref{tab:lee_statistic}. 

The main result here is that the correlation between $\mathbf{u}$ and $\mathbf{t}$ is low and non-significant (for any reasonable threshold) when measured with the Pearson coefficient. However it is highly significant, under both null models, when using $L$. Since $\mathbf{t}$ is just $\mathbf{u}$ with some additional steps of value propagation elsewhere in the network, we would expect some correlation between the datasets, which is captured by $L$ but not $\rho$. This is because the random noise component is quite strong, and it requires the neighbour averaging performed by $L$ to make the `hidden' correlation visible. The correlation between $\mathbf{v}$ and $\mathbf{t}$ is significant under both metrics, however the absolute value of $L$ is quite low. Again this seems to be a better reflection of the data, since in three of the four communities the data are only very weakly related and there is a fairly strong noise component in the fourth.

\section{Correlograms}\label{sec:correlogram}

\subsection{Definition}
The last concept we introduce is the \emph{network correlogram}, which can be constructed by making the weight matrix a function of `distance class'. In the case where $W$ is the adjacency matrix there is a quite natural mapping where $w_{ij}(d)=1$ if the shortest path between $i$ and $j$ has length $d$ otherwise the elements are zero. Then for any statistic the corresponding correlogram is obtained by substituting $W(d)$ into the definition in place of $W$. For example, the Moran correlogram is
\begin{equation}
    I(d) = \frac{N}{|W(d)|} \frac{ \sum_{ij}^N (x_i - \bar{x}) w_{ij}(d) (x_j - \bar{x}) }{ \sum_i^N (x_i - \bar{x})^2}
\end{equation}
Statistical significance is assessed using a data-permutation or configuration model null. Since there is a test for every value of $d$, again one must be wary of the multiple comparisons problem and the same approaches as for the local indices (low $\alpha$, Bonferroni Correction, False Discovery Rate) can be implemented, though again, uncritical application of statistical significance is not recommended and significance should be interpreted as an indication of potentially interesting behaviour rather than a definitive `finding'.

\subsection{Example}

\begin{figure}
    \centering
    \includegraphics[width=\textwidth]{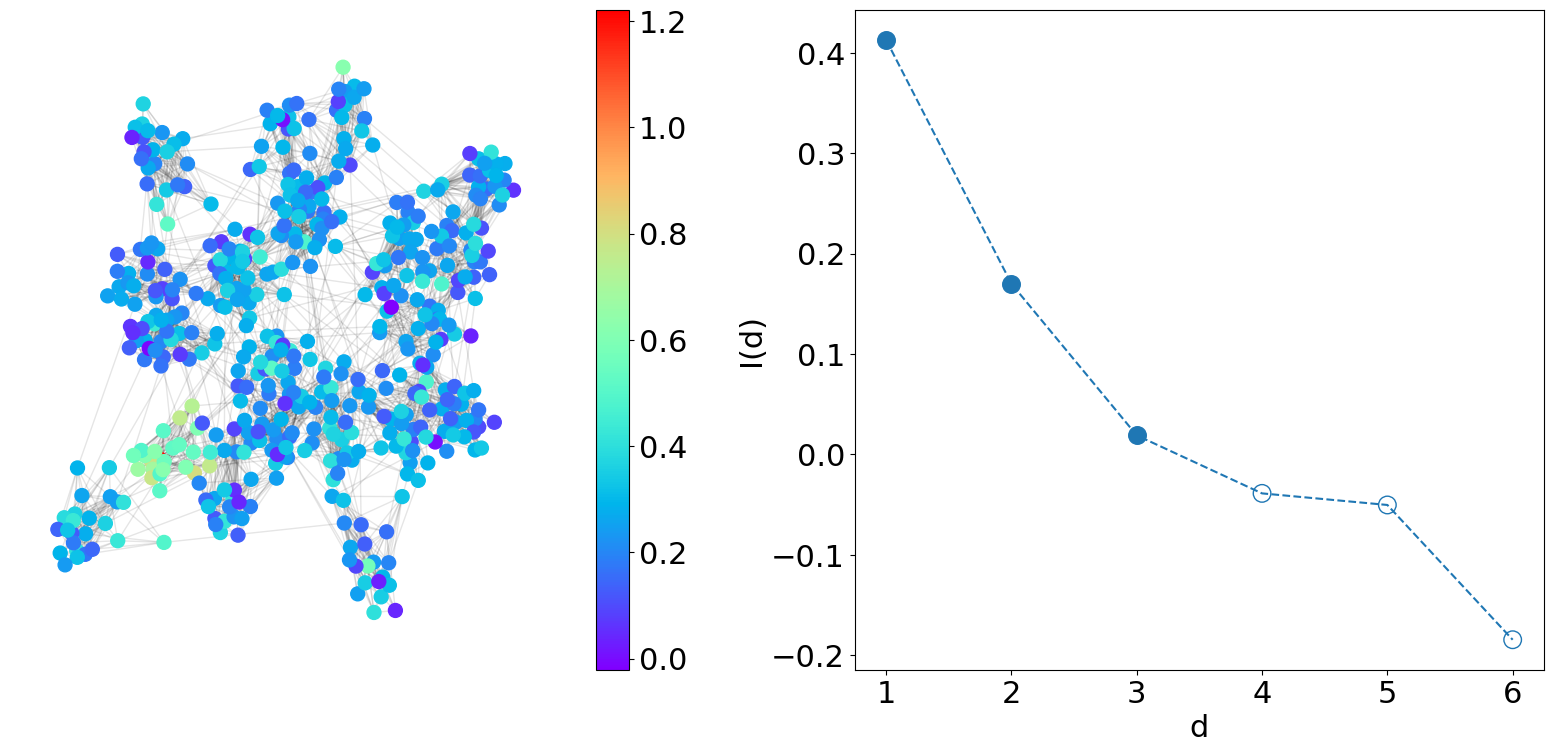}
    \caption{Left: LFR network. Right: Moran Correlogram $I(d)$. Solid blue points have p-value (under the data permutation null) $<0.01$, open points are above this threshold.}
    \label{fig:correlogram}
\end{figure}
 Figure \ref{fig:correlogram} shows an LFR network with 500 nodes. Value propagation is run for $M=250$ steps with $\sigma=0.1$. The Moran correlogram, $I(d)$, decays as a function of distance class. The modular structure of the network means that nodes separated by paths longer than 2 have very weak clustering of node values, because the value propagation is largely restricted to the source module. $I(d>3)$ is negative but the values are not statistically significant. Note also that the average shortest path in the network is $\simeq 3.7$, so there are fewer and fewer edges in the weight matrix for $d > 3$ and the largest values of $d$ have many isolated nodes, $\sum_j w(d)_{ij} = 0$.

\section{Wikipedia Network Analysis}\label{sec:wikipedia}

The online user-contributed encyclopedia Wikipedia has long been a subject of study for network science \cite{bellomi2005network,zlatic2006wikipedias}, with the aim to understand how relationships between different topics are encoded in the network of hyperlinks between pages. In this section we will analyse a portion of the Wikipedia network. Although of some intrinsic interest, the main purpose is to show how the methods described above can be applied in the exploratory data analysis of a realistic dataset.

We construct the \emph{EgoMinusEgo} network \cite{coscia2012demon} of the Wikipedia page for \emph{Network science}\footnote{\url{https://en.wikipedia.org/wiki/Network_science} Accessed 03/05/24} using MediaWiki Links API\footnote{\url{https://www.mediawiki.org/wiki/API:Links} Accessed 03/05/24}. This means:
\begin{itemize}
    \item We obtain all the outgoing links from the \emph{Network science} page ($N=433$ at time of access) to get the list of pages of interest $V$.
    \item For every page in $V$ we find all of its links.
    \item The network is constructed by considering only links between the nodes in $V$.
\end{itemize}
Though hyperlinks are intrinsically directional, for simplicity we consider an undirected network, so for a hyperlink between pages $i$ and $j$ in either direction we add a link $ij$. We also use the row normalised adjacency matrix as the weighting matrix $W$.

There are a number of possible data associated with a Wikipedia page: page length, number of edits, number of watchers, length of the corresponding Talk page etc. We will primarily consider the number of page views. Using the Massviews API\footnote{\url{https://pageviews.wmcloud.org/massviews/} Accessed 03/05/24} we collect the number of views for all pages connected to \emph{Network science} in the month of April 2024. We remove one page with 0 page views, this is from a link to a page that has not been created. Because of the huge range of page views spanning from over 3 million for the most popular page\footnote{\url{https://en.wikipedia.org/wiki/YouTube}  Accessed 03/05/24} to just 4 for the least popular\footnote{\url{https://en.wikipedia.org/wiki/Search_engine_spammer} Accessed 03/05/24} we use the (base 10) log of the view count as the node data. 

\begin{figure}
    \centering
    \includegraphics[width=\textwidth]{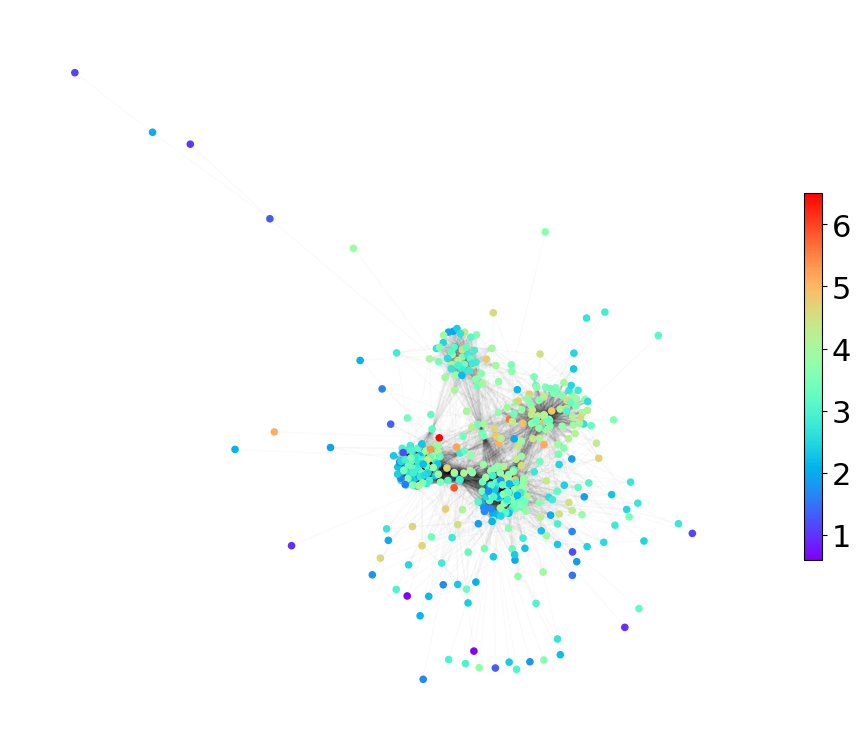}
    \caption{\emph{Network science} EgoMinusEgo network. Nodes coloured by log of page views.}
    \label{fig:wikipage}
\end{figure}
\begin{figure}
    \centering
    \includegraphics[width=\textwidth]{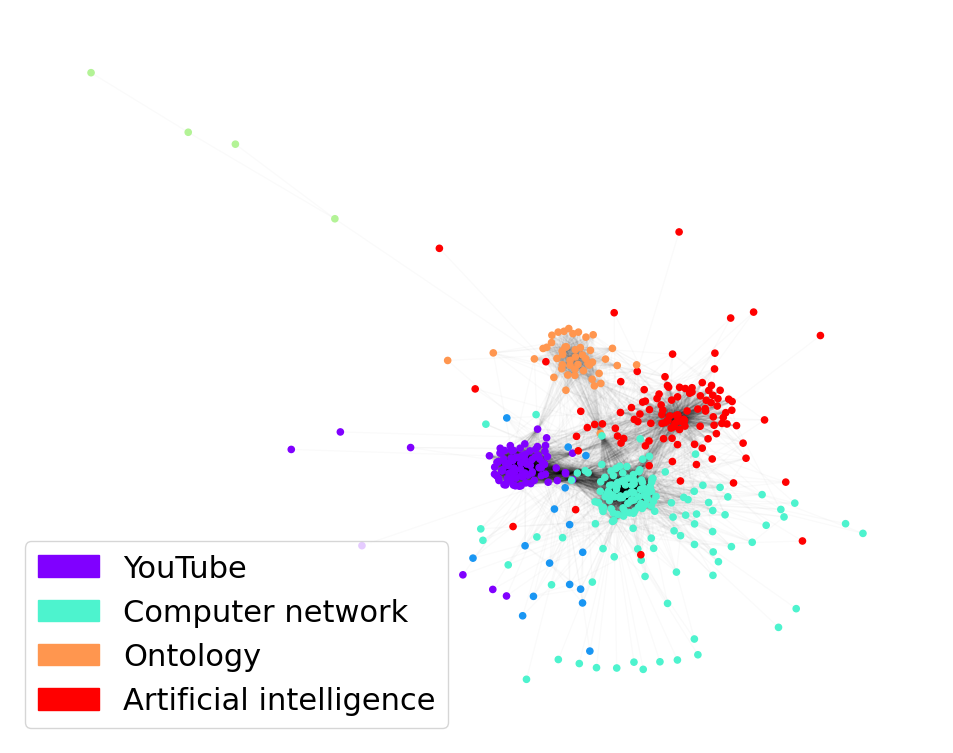}
    \caption{\emph{Network science} EgoMinusEgo network. Large communities detected by Louvain algorithm.}
    \label{fig:wikicomm}
\end{figure}
Figure \ref{fig:wikipage} shows the network with the nodes coloured by page views. Community detection is straightforward, and the Louvain algorithm \cite{blondel2008fast} identifies the 4 communities implied by the force directed layout and two small communities consisting of the outlying nodes which we will largely ignore. Modularity $Q=0.55$. The communities shown in Figure \ref{fig:wikicomm} are labelled by their most viewed page. Broadly: the \emph{YouTube} community consists of pages about social networks, the \emph{Computer network} community has technical material about network science, the \emph{Ontology} community has pages about organising information and the \emph{Artificial intelligence} community consists of a number of pages  about topics where network science methods have been applied.

\begin{figure}
    \centering
    \includegraphics[width=\textwidth]{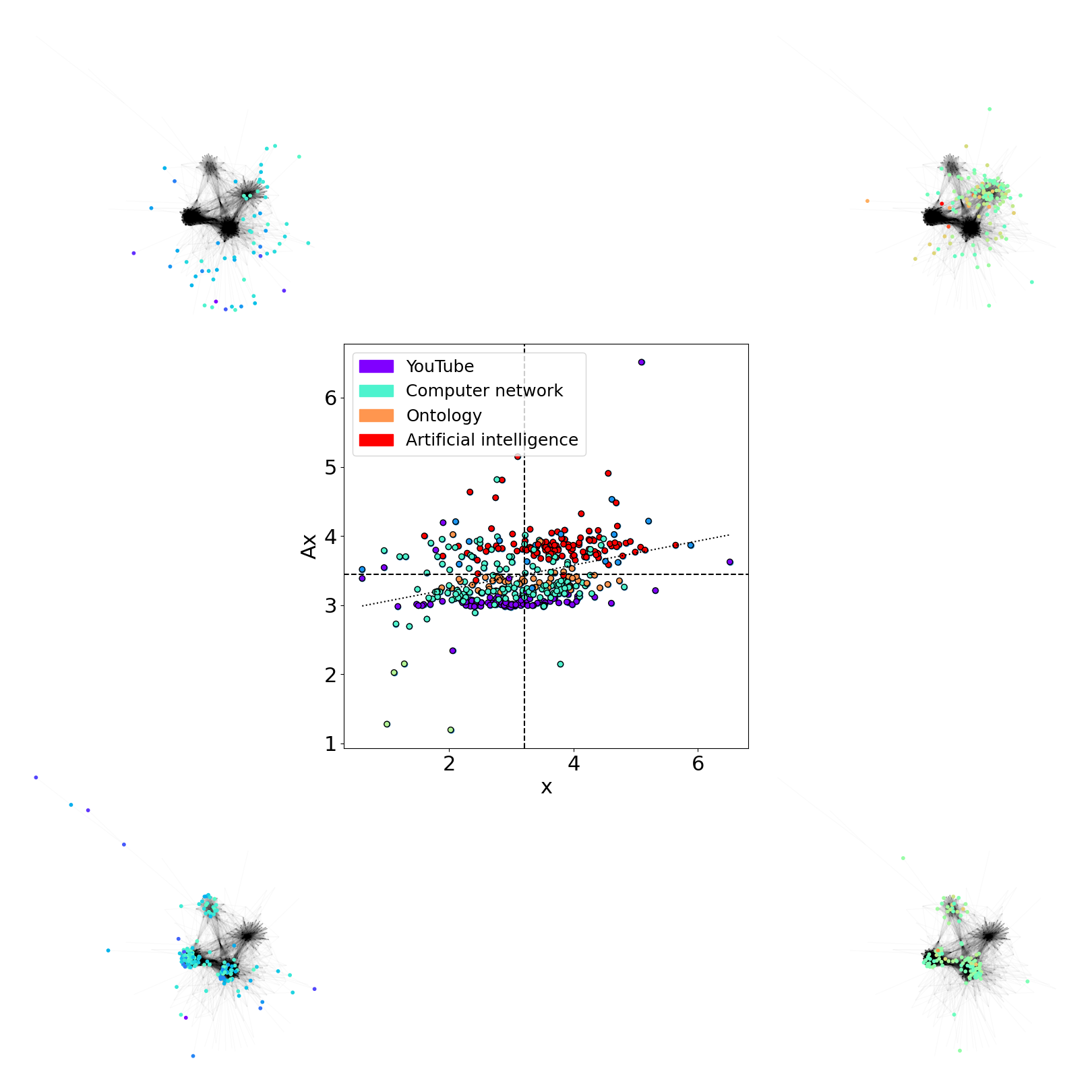}
    \caption{Moran scatter plot for \emph{Network science} EgoMinusEgo network. Nodes in each quadrant are shown in the inset network figures. The slope of the dotted line equals Moran's $I$.}
    \label{fig:wikiscatter}
\end{figure}
The network Moran index of the page view data is
\[
I = 0.174
\]
with vanishing p-value under both null models. This indicates weak but significant clustering of page views. We will set the significance threshold to $\alpha=0.01$ throughout, but again note we are not advocating a standard hypothesis testing framework, rather we use this threshold only to highlight `interesting' data in this exploratory analysis. The correlogram (not shown) indicates significant, but very small, autocorrelation at $d=2$, $I(2) = 0.029$ while $I(d\geq3)$ are not significant. The average shortest path is $\simeq 2.3$ and for $d\geq3$ we have fewer and fewer edges in $W(d)$. This suggests that page views are not clustered using next-nearest neighbours and beyond so we can restrict our analysis to nearest neighbours.

The Moran scatter plot is shown in Figure \ref{fig:wikiscatter}. This indicates that the \emph{Artificial Intelligence} community is a high-high cluster, popular pages connected to other popular pages. The other 3 communities are split between low-low and high-low, suggesting a mixture of popular and unpopular pages in the \emph{YouTube}, \emph{Ontology} and \emph{Computer Network} communities. There are a small number of low-high pages i.e. unpopular pages connected to popular ones. These are not associated with one particular community and consist of a number of biographies and relatively obscure topics of marginal relevance to Network science, which only touch on some of the central topics in the subject.

\begin{figure}  
    \centering
    \includegraphics[width=\textwidth]{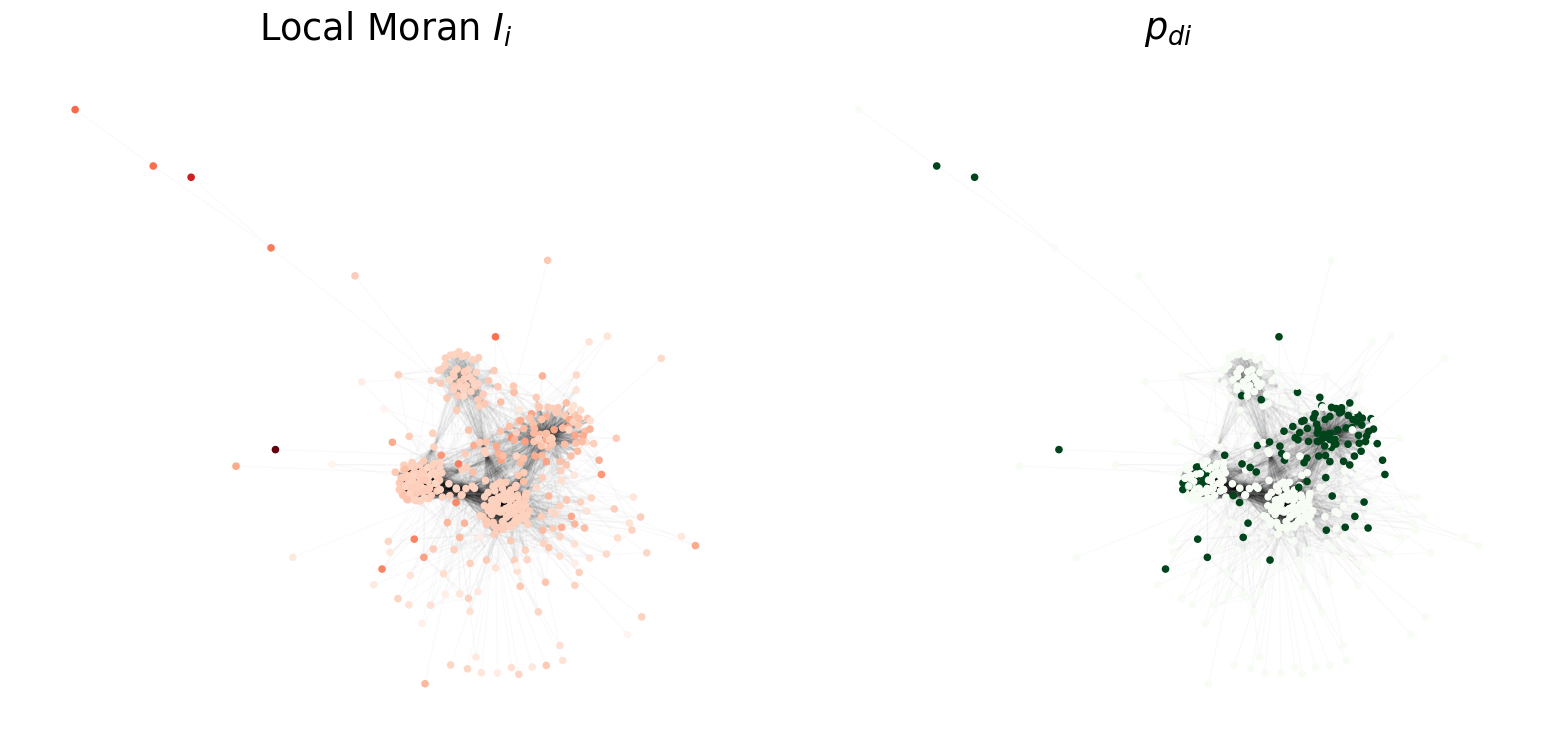}
    \caption{Left: Node colour indicating $I_i$, node Moran value (darker red is higher). Right: Node color indicating p-value of $I_i$ under the data permutation null (configuration null is quite similar). Values less than $0.01$ are dark green, higher values are very light green.}
    \label{fig:wikilocal}
\end{figure}
The analysis of the Moran scatter plot is complemented by plotting the node Moran index, as shown in Figure \ref{fig:wikilocal}. While there is no obvious pattern in the value of $I_i$, when assessing the significance under either null model many nodes in the \emph{Artificial Intelligence} community are significant, indicating clustering of (high) page views in that community. Here, popular topics are preferentially connected, which is not the case for the other communities.

\begin{figure}
    \centering
    \includegraphics[height=0.95\textheight]{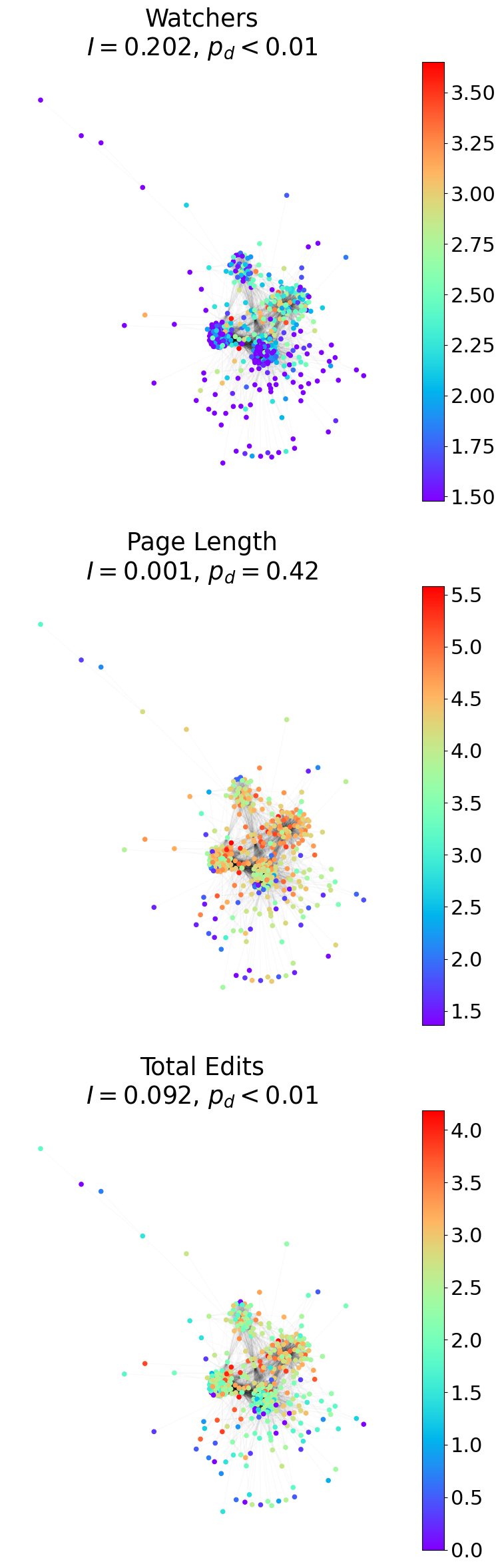}
    \caption{Same network with nodes coloured according to other data values.}
    \label{fig:wikiother}
\end{figure}
Figure \ref{fig:wikiother} shows a number of other statistics for each page in the network obtained from the corresponding `info' page\footnote{e.g. \url{https://en.wikipedia.org/wiki/Network_science?action=info} Accessed 03/05/24}. The three quantities shown are:
\begin{itemize}
    \item The number of \emph{watchers}, that is, Wikipedia users who have chosen to be notified when that page is edited.
    \item The page \emph{length}, in bytes.
    \item The total number of \emph{edits} made to the page since its creation.
\end{itemize}
As with page views these quantities span a very large range, so we have logged the values. While \emph{edits} and \emph{watchers} are autocorrelated, indicating clustering, \emph{length} is not, implying that we cannot predict the length of a page based on the length of its neighbours.

\begin{table}
    \centering
    \begin{tabular}{|c||c|c||}
    \hline
          & $\rho$ & $L$\\ \hline
\emph{Page views},\emph{watchers} & 0.87 & 0.27 \\ \hline
\emph{Page views},\emph{length}& 0.67 & 0.17 \\ \hline
\emph{Page views},\emph{edits} & 0.80 & 0.22 \\ \hline
    \end{tabular}
    \caption{Pearson and Lee correlation of page views with other page metrics. All values are significant at $\alpha = 0.01$. Note the $L$ statistic is computed with 1s on the diagonal of the adjacency matrix as mentioned in Section \ref{sec:bivariate}. }
    \label{tab:viewcorrelation}
\end{table}
Table \ref{tab:viewcorrelation} shows the Pearson and Lee correlation between the page view data and the data from Figure \ref{fig:wikiother}. All Pearson and Lee correlations are significant here, however the values are quite different. Generally autocorrelation of either or both datasets inflates the Pearson correlation and the value and quoted significance cannot be trusted see \cite{clifford1989assessing} or more recent work in \cite{arthur2024general}. The Lee correlation mitigates this and indicates a modest degree of correlation between page views and the other variables.

\begin{figure}
    \centering
    \includegraphics[width=\textwidth]{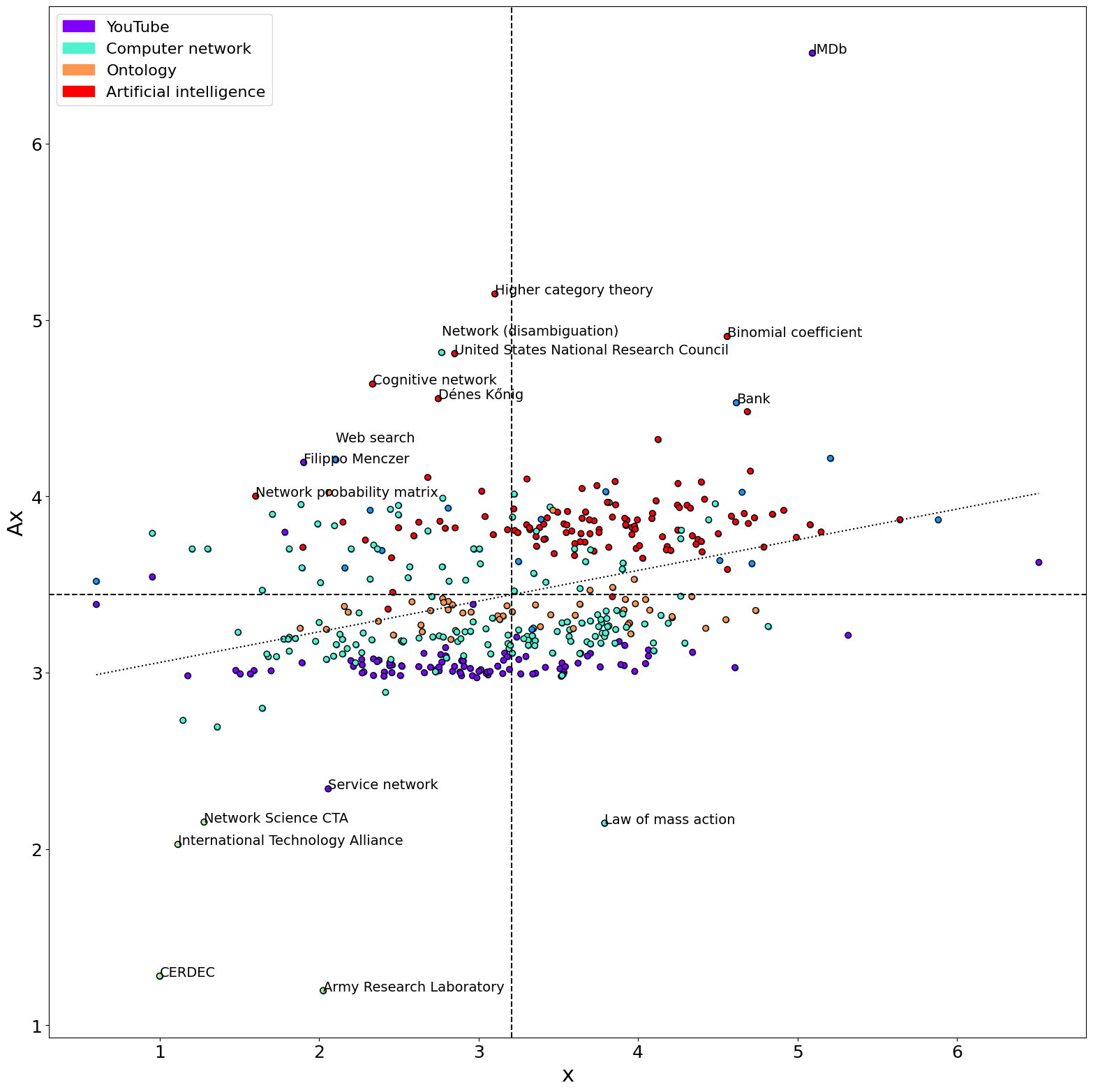}
    \caption{Moran scatter plot for page view data with outliers highlighted.}
    \label{fig:wikioutlier}
\end{figure}
Finally, Figure \ref{fig:wikioutlier} shows the Moran scatter plot for page views once more, this time with outliers highlighted. Some examples here are illuminating:
\begin{itemize}
    \item A high-high outlier is the article for the Internet Movie Database \emph{IMDb}. The IMDb is a key example in many network science papers and books. However the IMDb article does not discuss network science, but does mention YouTube, the most viewed page in the network.
    \item A low-high outlier is the article  \emph{Cognitive network}. This mentions a few general, high-traffic topics like \emph{Artificial intelligence} and \emph{Network management}. This is an article about computer science more than network science and so is only linked to \emph{Network science} via general, popular articles.
    \item A low-low outlier is the article about the \emph{Army Research Laboratory}. This is a low traffic page that is only connected to a couple of other topics related to this laboratory.
    \item A high-low outlier is the page for the \emph{Law of Mass Action}, a relatively popular page about a principle in chemistry that is only marginally related to network science.
\end{itemize}
In different applications, or with different data, the importance and interpretation of outliers might be different. Here all the outliers seem to be quite peripheral topics, perhaps not worth including in the main article about network science. For application to Wikipedia, this kind of analysis could be used to detect inappropriate links or malicious edits \cite{kharazian2023governance}. Looking at the other variables could give interesting insights for editors, for example anomalous page lengths could indicate articles in need to improvement, but we leave this for future work.

\section{Discussion}

This paper shows how spatial summary statistics and methods of exploratory spatial data analysis can be applied to complex networks. We have focused on Moran-type statistics, which use dot products of the data  with the neighbour averaged data to give global and local measures of data clustering, and related  tools like the Moran scatter plot and correlogram.

We have presented a number of illustrative examples on real and synthetic data. These show that data clustering on networks is related to but distinct from community structure. The Wikipedia network had four strong communities but page views were only significantly clustered in one of them, and page length was not clustered at all. We also suggest that correlation between different data observed on the same network should not be measured with Pearson correlation. This is for essentially the same reason that causes problems in spatial and time series analysis - value and significance inflation due to autocorrelation \cite{clifford1989assessing}. We suggest Lee correlation instead, which seems to perform better in accounting for autocorrelation.

The extension of the above to weighted networks or other distance functions is obvious, consisting of the replacement of $w_{ij}$ with the corresponding matrix. The values of the various statistics will change substantially and thus interpretation might not be as straightforward, however the basic framework of significance testing introduced will still be valid. In that respect the most novel feature of the present work is the applying the configuration model to these statistics. For a spatial network this is less likely to be useful, but for a network of social contacts or hyperlinks, looking at summary statistics in the context of changing network structure seems very apt.

We note that there are a large number of spatial statistics with a similar purpose to Moran's $I$. The most prominent are probably Geary's C \cite{anselin2019local} and Getis-Ord statistics \cite{getis1992analysis}. Similar analyses could be performed with these statistics, though we omit this for the sake of brevity. Statisticians have also expended much effort on analytical calculations of the variance of Moran's I and other spatial statistics \cite{cliff1981spatial}. With modern computers, permutation and resampling methods are preferred \cite{anselin2009geoda}, however for very large networks sampling from the configuration model or permuting data (especially the conditional permutation required for the node Moran index) is quite time consuming and analytical approximations would be useful.

We hope that researchers studying data on complex networks find this exposition interesting and can use these statistics in their own work. Implementations of all the Moran-type statistics discussed here, and more, are available in spatial analysis tool kits like GeoDa \cite{anselin2009geoda} and PySAL\cite{rey2009pysal}. However, for the benefit of practitioners working within common software frameworks for complex networks we offer implementations in Python
\url{https://github.com/rudyarthur/network_correlation} using the NetworkX library \cite{hagberg2008exploring}. We include Geary's C, Getis-Ord statistics and a number of others in this package for completeness.

\bibliographystyle{plain} 
\bibliography{sample}

\end{document}